\documentclass[conference]{IEEEtran}
\IEEEoverridecommandlockouts
\usepackage{cite}
\usepackage{amsmath,amssymb,amsfonts}
\usepackage{algorithmic}
\usepackage{graphicx}
\usepackage{textcomp}
\usepackage{xcolor}
\usepackage{url}

\newcommand{\parhead}[1]{\vspace{0.5pt plus 2pt minus 1pt}\par\noindent\textbf{#1.}\hspace{1em plus 0.5em minus 0.5em}}

\def\BibTeX{{\rm B\kern-.05em{\sc i\kern-.025em b}\kern-.08em
    T\kern-.1667em\lower.7ex\hbox{E}\kern-.125emX}}
\begin{document}

\title{Writing a Good Security Paper for ISSCC (2025)
}

\makeatletter 
\newcommand{\linebreakand}{%
  \end{@IEEEauthorhalign}
  \hfill\mbox{}\par
  \mbox{}\hfill\begin{@IEEEauthorhalign}
}
\makeatother 

\author{
\IEEEauthorblockN{Utsav Banerjee}
\IEEEauthorblockA{\textit{Indian Institute of Science} \\
Bengaluru, India \\
utsav@iisc.ac.in}
\and
\IEEEauthorblockN{Chiraag Juvekar}
\IEEEauthorblockA{\textit{Apple} \\
San Carlos, CA \\
chiraag@apple.com}
\and
\IEEEauthorblockN{Yong Ki Lee}
\IEEEauthorblockA{\textit{Samsung Electronics} \\
Suwon, Korea \\
yongki93.lee@samsung.com}
\linebreakand
\IEEEauthorblockN{Leibo Liu}
\IEEEauthorblockA{\textit{Tsinghua University} \\
Beijing, China \\
liulb@tsinghua.edu.cn}
\and
\IEEEauthorblockN{Sanu Mathew}
\IEEEauthorblockA{\textit{Intel} \\
Hillsboro, OR \\
sanu.k.mathew@intel.com}
\and
\IEEEauthorblockN{Thomas Poeppelmann}
\IEEEauthorblockA{\textit{Infineon Technologies} \\
Neubiberg, Germany \\
thomas.poeppelmann@infineon.com}
\and
\IEEEauthorblockN{Shreyas Sen}
\IEEEauthorblockA{\textit{Purdue University} \\
West Lafayette, IN \\
shreyas@purdue.edu}
\linebreakand
\IEEEauthorblockN{Takeshi Sugawara}
\IEEEauthorblockA{\textit{The Univ. Electro-Communications} \\
Tokyo, Japan \\
sugawara@uec.ac.jp}
\and
\IEEEauthorblockN{Ingrid Verbauwhede}
\IEEEauthorblockA{\textit{KU Leuven} \\
Leuven, Belgium \\
ingrid.verbauwhede@kuleuven.be}
\and
\IEEEauthorblockN{Rabia Tugce Yazicigil}
\IEEEauthorblockA{\textit{Boston University} \\
Boston, MA \\
rty@bu.edu}
}

\maketitle

\section{Introduction}

Security is increasingly more important in designing chips and systems based on them, and the International Solid-State Circuits Conference (ISSCC), the leading conference for presenting advances in solid-state circuits and semiconductor technology, is committed to hardware security by establishing the security subcommittee since 2024. 
In the past two years, the authors of this paper reviewed submissions as members of the Security Subcommittee, a part of International Technical Program Committee (ITPC). This paper aims to encourage high-quality submissions to grow this field in the overall scope of the ISSCC.

The first purpose of the document is to share our knowledge in writing good security papers.
Hardware security is an interdisciplinary field, and writing papers in this area requires special attention, and we observed common and repeated mistakes. 
Although we are making an effort to improve the situation by providing review feedback to authors, the change made in this way occurs only gradually, and this paper is intended to accelerate the process.
The first two sections are devoted to providing insight into how we will review and rate papers and how to avoid common mistakes.
Section~\ref{sect:general} begins with the general hints applicable to most papers. 
In Section~\ref{sect:subcat}, we define specific requirements for papers that fall into specific subcategories, namely (i) implementations of fully homomorphic encryption (FHE) and post-quantum cryptography (PQC), (ii) evaluation of physically unclonable function (PUFs) and random number generator (RNG), and (iii) countermeasures against side-channel attacks (SCAs) and fault-injection attacks (FIAs).

The second purpose of this document is to expand the scope of potential submissions.
In the past two years, many papers have been accepted from the aforementioned three categories (i)--(iii).
Researchers typically select conferences for submission based on previously accepted papers, and we are afraid of a situation where only similar types of papers continue to be submitted.
We are open to more diverse papers, and Section~\ref{sect:newfield} describes the other categories that we would like to see submitted in the future.

\subsection{Disclaimer}
This document captures the perspectives of the authors as of May 2025, noting that these opinions may evolve with time or community input. It compiles the personal viewpoints of the members and does not represent the official stance of their respective organizations or the ISSCC in general.

\section{General Hints} \label{sect:general}

This part provides general recommendations likely to help most security papers. For additional tips, the readers may also consult the guide by Van der Spiegel and Smith~\cite{b1}.

\subsection{Innovation and Significance}

\textit{Innovation} and \textit{Significance} as defined below are the two main criteria in  reviewing papers~\cite{keith}:

\parhead{Innovation}
The conference attendees want to see new ideas leading to new research directions, and this criterion values new designs, methods, products, solutions, applications, or new problem to solve with immediate or potential value, such as novel circuits and system architectures.

\parhead{Significance}
Attendees also want to see the state-of-the-art, especially those from industry, and this criterion covers a positive impact on the system, product or SSCS community, including industry firsts, technology firsts, high-volume product deployment, and state-of-the-art advancements.

Although a paper with outstanding innovation and significance is ideal, the ISSCC ITPC aims to accept many papers with either outstanding innovation or outstanding significance. For example, although a highly innovative paper may not have the best figure-of-merit (FoM), these contributions explore new circuit and system architectures that may approach problems from a unique perspective, challenge fundamental tradeoffs, or even open up new directions for future research and products. In contrast, a highly significant paper, describing an industry first or state-of-the-art advancements in ICs and SoCs, may be based on prior works. When innovative ideas are eventually integrated into products and/or advance the field, this is an important validation of the underlying design techniques. ISSCC ITPC members are instructed to review each paper based on the innovation and significance of the work.

\subsection{Evaluation}

\parhead{Benchmark}
Papers on a circuit, building block, architecture, or system should provide comparisons with state-of-the-art publications on industry-recognized benchmarking methods and metrics.
These benchmarks and comparisons should be conducted in a similar level (e.g., system, chip, subsystem, etc.), and if not, these distinctions should be clearly indicated and clarified.

\parhead{Performance Details}
Circuit performance is affected by multiple factors, and some articles lack critical details to separate the improvement based on the claimed innovation from those based on other factors. 
Thus, make sure to provide enough information that the claimed innovation actually leads to an improvement in an important key performance indicator (KPI). 
Here are more concrete suggestions:

\begin{itemize}

\item Provide measurement conditions such as voltage, frequency, and temperature. 

\item Clarify what blocks or functions are included (or omitted) in the evaluation, especially for power consumption, area, and latency measurements. In case of omissions, e.g., off-chip memory accesses, these need to be explained sufficiently.

\item The evaluation of the area should include details such as the technology node and the circuit size in mm$^2$. 
For papers aiming at technology-independent innovation, e.g., computing architecture, technology-independent metrics should be provided, such as the number of gates normalized with a 2-way NAND (GEs), the number of FFs, the area and the capacity of SRAMs, and other hard macros used. 
If possible, provide a breakdown of the circuit into sub-blocks to allow for a detailed evaluation.

\item If runtime is measured, it should be provided in absolute numbers (e.g., milliseconds) and clock cycles. 

\item Provide sufficient details on power consumption, especially when it is claimed as the main KPI.
Discriminate active and passive power consumption and provide information on peak power consumption.

\item Provide measurements across different voltage and frequency settings, especially when the chip claims a wide operating condition or features dynamic voltage and frequency scaling.

\end{itemize}

\parhead{Clarification of Silicon Boundary}
The measured and simulated results should be clearly discriminated. 
In general, the main results should be based on real measurements, and simulation should be limited for alternative evaluations, such as (i) covering scenarios that are not implemented, (ii) initial design-space exploration, and (iii) presilicon validation that is later compared with real measurement.

\parhead{Out-of-Scope Components}
The essence of security design can lie in the details, but authors can omit or simplify components that are not the main focus of the paper.
For example, a key register that allows external read and write access is an immediate security problem, but such a design is acceptable unless such access control is the main subject of a paper. 
An LFSR is unacceptable as a cryptographic random number generator, but it can be used for some evaluation as long as the RNG is not the main technical contribution. 
Meanwhile, we should avoid readers from replicating the simplification for practical development, which can cause a security incident. 
Therefore, such non-ideal aspects should be clearly described as limitations in the paper.

\subsection{Research Direction}

\parhead{Process Node}
Papers that explore the cutting edge in terms of process node are exciting. However, the submission of papers in mature process nodes is still highly encouraged. Such papers make sense when architectural innovation is targeted and when no specific properties of a process node are used or required. For such papers, make sure to provide sufficient information on the architectural innovation or process-independent circuit-level techniques used.

\parhead{Compliance to Standards of Cryptographic Security}
The security of cryptography is either rigorously proved or evaluated. 
Public-key schemes, including PQC and FHE, commonly have provable security reduction to mathematical problems that are believed to be hard, such as learning with errors. 
Similarly, formally provable security is becoming common in masking schemes used for SCA protection. 
Meanwhile, the security of other cryptographic primitives, such as the advanced encryption standard (AES), is supported by rigorous security evaluation with several cryptanalytic techniques. 
Papers on realizing a cryptographic scheme should cite either a standard document~\cite{barker} or previous papers justifying the scheme's cryptographic security.
Making disruptive changes to existing schemes, which invalidates their existing security proofs and evaluation, is unacceptable.

\parhead{Silicon Realization of Previous Techniques}
We are open to papers about silicon realization of techniques previously published in other venues (e.g., IACR Transactions on Cryptographic Hardware and Embedded Systems (TCHES)) without silicon.
The authors are encouraged to put a particular focus on the circuit techniques that are used to realize the silicon and how they affect the outcome (e.g., compare a straightforward with a circuit-level optimized implementation).
In such a case, the previous work should be cited and described in the paper, and we appreciate a clear statement if a paper mainly targets the circuit-level realization. 
Papers that try to hide this relationship or make unclear claims about innovation may be rejected. 

\subsection{Writing}

\parhead{Clear and Reasonable Motivation}
The introduction of the paper should be concise and logical, and the motivation and technical innovation should be consistent.
For example, low-cost battery-powered IoT devices are inappropriate targets to motivate a chip with a huge area ($>$10mm$^2$), a state-of-the-art process node, significant power consumption, and more cryptographic performance than necessary. 
Meanwhile, papers targeting a very narrow and niche application will be ignored, because only few audience at ISSCC will be interested in them.

\parhead{Avoid Describing Well-Known Standards or Common Knowledge}
ISSCC papers need to provide a lot of information on few pages. 
Do not waste your space by describing well-known standards or common knowledge. For example, when implementing AES, citation to an external document~\cite{aes} is sufficient, and it is not necessary to explain how AES works in general. 
Instead, use the space to describe the actual security aspects and the details of your contribution and how it drives innovation and significance.

\subsection{Collaboration}
In addition to the tips in this paper, we encourage silicon design groups to team up with security researchers specialized in attacking silicon that may lack the ability to design silicon chips.
The same applies to groups working on tools.

\section{Requirements for Particular Sub-Categories.} \label{sect:subcat}

\subsection{Cryptographic Accelerators}

We believe that constant execution time is a baseline requirement for cryptographic accelerators, including FHE and PQC implementations, because there are several attacks that remotely exploit the timing side channel. 
Please provide a statement if a cryptographic hardware accelerator runs in constant time (i.e., execution time independent of any secret input). In case the device is non-constant time, it should be explained why that is acceptable. 

\subsection{RNG}

\parhead{Discrimination of Noise Source and Post-Processing}
Discriminate the circuit technique for harvesting the entropy from physical random noise and cryptographic post-processing.
The first half, circuit techniques for harvesting entropy, is usually more interesting.
When a random noise source is combined with post-processing, then the description should still allow an independent evaluation.
Evaluation of bitstream after post-processing is insufficient to validate the quality of noise source; provide
the results of the noise source itself, for example, by using NIST standard randomness test~\cite{sp800-22}.

\parhead{Stochastic model for TRNG}
Provide a stochastic model that justifies the rate of entropy that your circuit generates. 
In general, TRNG evaluation should follow the notions described in NIST SP 800-90B~\cite{sp800-90b} and AIS31~\cite{ais} as closely as possible.

\parhead{System performance}
While high-performance noise sources are certainly of interest, we are also open to designs considering practical design trade-offs for particular applications. 
For example, low-cost IoT devices are unlikely to need a high bitrate, and a minimal noise source, combined with a deterministic RNG to expand the available bitstream, can be a practical research challenge.

\subsection{PUF}

\parhead{Stochastic Model}
Similar to RNG, a PUF design should be accompanied by a stochastic model.

\parhead{Strong PUF}
Strong PUF claims need to be substantiated. 
We do not encourage ad-hoc strong PUF submissions that will be broken soon after. 
When you propose a strong PUF, you should provide a rationale why modeling attacks are hard, along with experimental evidence that the proposed withstand state-of-the-art attacks.

\parhead{Correlation}
The authors should evaluate the independence between the response bits generated through PUFs with appropriate statistical tests, such as those included in NIST SP800-22~\cite{sp800-22}.
Even if PUF cells look independent, they can generate correlated responses by several reasons, including coupling in power delivery network and a circuit component shared among multiple PUF cells.

\subsection{Side-channel attacks}

\parhead{Attack evaluation}
An unsuccessful attack does not validate your defense if the attack is poorly conducted, and the paper should provide evidence that the attack is reasonably executed. 
One common way is to show the results with countermeasures turned on versus off, and to use a successful attack with the countermeasure turned off as a baseline.
You should use standard tests (e.g., correlation power analysis (CPA) and test vector leakage assessment (TVLA)) and provide details of measurement setup. 
Since even an affordable oscilloscopes can easily acquire more than 1 million traces, the evaluation is considered insufficient when it uses much fewer number of traces.
At the same time, a circuit-level security technique broken with a large number of traces is acceptable, and we encourage authors to be open about this. 
In such a case, the evaluation of the cost vs. benefit of the technique and the composability with other techniques (e.g., masking) should be in the focus of the work.

\parhead{Attacker model}
The paper should claim the threat that the proposed method is addressing, along with experimental results to support the claim. 
We are open to papers focusing on a particular threat, i.e., targeting SCAs while excluding FIAs.
If a design implements multiple countermeasures to address different threats, experimental results should be included for each individual technique, as well as their combined effect.

\parhead{Masking and Threshold Implementation}
We recommend using masking schemes with formal security proof over the old ones with ad hoc security only.
For masking implementations, provide the experimental results with RNG on and off to provide the baseline. 
Attack results beyond the theoretical guarantee of the masking scheme, e.g., 2nd-order attack on 1st-order masking, are appreciated.
Make sure that the assumption behind masking is satisfied; Masking schemes without glitch (i.e., dynamic hazard) robustness cannot be used without additional measures to prevent glitches.

\subsection{Attack Sensors}

\parhead{Target Attacks}
Papers on on-chip sensors for detecting attacks should claim the target attacks, e.g., clock glitching and laser fault injection (LFI), along with experimental results to support the claim. 

\parhead{Attack Parameters}
With FIAs, an attacker can typically explore the parameter space, e.g., pulse patterns in clock glitching and laser-spot coordinate in LFI.
The paper should argue that the evaluation covers the reasonable attack space and that the critical case is not excluded. 
One way to achieve this is an exhaustive evaluation by sweeping the parameters. 
Another way is to choose worst-case attack parameters considering a white-box analysis, i.e., the attacker knows the details about the defense and tries to evade it. 

\parhead{Attacks on Sensors and Denial of Service}
Ironically, sensors for detecting attacks can be a new target of attack.
In particular, frequent false positives lead to a reliability problem, 
and an attacker may even intentionally activate the sensors to cause a denial of service. 
Provide a discussion on how to prevent or mitigate such attacks.

\section{Field and papers we would like to see submitted to the security track} \label{sect:newfield}

Popular research topics in the past two years include FHE/PQC accelerators, PUF, RNG, and SCA and FIA countermeasures, but we are open to more diverse topics on hardware security. 


\parhead{Non-Cryptographic Targets}
Hardware attacks on non-cryptographic targets are emerging, and we like to see defenses against those attacks. 
Potential targets include:
\begin{itemize}
\item Machine-learning accelerators
\item Analog-to-digital converters
\item (General) sensors, such as image sensors, LiDARs, and microphones
\end{itemize}

\parhead{Secure Processor Designs}
Processors and microcontrollers with certain security features, such as:
\begin{itemize}
\item Microarchitecture with resistance against fault attacks, e.g., redundant cores and lockstepping
\item Defenses against microarchitectural SCA (e.g., cache attacks and cold-boot attacks) and FIA (e.g., row-hammer attacks)
\item Microarchitectural support for secure software execution, e.g., secure enclave, memory encryption, and control-flow integrity 
\end{itemize}

\parhead{System Integration and Applications}
We are open to papers that focus on the integration of several secure primitives into a system, especially those with a particular application.
\begin{itemize}
\item Secure SoC design, including cryptographic accelerators, hardware-based isolation mechanisms, and secure boot.
\item Integration of several countermeasures based on a concrete threat model.
\item Security in constrained environments, e.g., realizing security on tiny sensors or microcontrollers with minimal circuit area and power budget. 
\item Security on power- and energy-constrained devices, e.g., NFC/smart cards that should work with wireless power supply. 
\item Security in a large-scale systems, e.g., multi-node clusters operated in data centers. 
\item Co-design of security and safety, e.g., automotive safety features as described in ISO26262. 
\end{itemize}

\parhead{Circuit Techniques for Provable Security}
We welcome papers that aim to bridge the gap between provable methods, such as masking, and circuit-level methods, such as equalizing power regulators. An approach involves combining provable techniques with hardware-level defenses to enhance security and efficiency. Another approach may focus on developing circuit-level methods and components to satisfy the prerequisites on which provable methods are based.

\parhead{Design Methodology}
Predicting and estimating security before fabrication is a challenging issue, especially when analog-domain methods are included. 
Thus, we are open to papers that showcase the design methodology for verifying security before fabrication and ensuring the predicted security in real silicon.

\section*{Acknowledgment} 
The authors wish to thank Keith Bowman and Marian Verhelst, the program chair and the vice chair of ISSCC~2026, for their valuable feedback to earlier version of this article.

\end{document}